\documentclass{elsart}
\usepackage{epsfig}
\usepackage[english]{babel}

\begin{document}
\begin{frontmatter}

\title{Local and cluster critical dynamics of the 3d random-site Ising model}
\author[Franko]{D. Ivaneyko\corauthref{cor}},
\corauth[cor]{Corresponding author.}
\ead{ivaneiko@ktf.franko.lviv.ua}
\author[ICMP]{J. Ilnytskyi},
\ead{iln@icmp.lviv.ua}
\author[Nancy]{B. Berche},
\ead{berche@lpm.u-nancy.fr}
\author[ICMP,Linz,Franko]{Yu. Holovatch}
\ead{hol@icmp.lviv.ua}

\address[Franko]{Ivan Franko National University of Lviv,
               79005 Lviv, Ukraine}
\address[ICMP]{Institute for Condensed Matter Physics,
               National Acad. Sci. of Ukraine,
               79011 Lviv, Ukraine}
\address[Nancy]{ Laboratoire de Physique des Mat\'eriaux,
               Universit\'e Henri Poincar\'e, Nancy 1,
               54506 Vand\oe uvre les Nancy Cedex, France}
\address[Linz]{Institut f\"ur Theoretische Physik,
               Johannes Kepler Universit\"at Linz,
               4040 Linz, Austria}

\begin{abstract}
We present the results of Monte Carlo simulations for the critical
dynamics of the three-dimensional site-diluted quenched Ising
model. Three different dynamics are considered, these correspond
to the local update Metropolis scheme as well as to the
Swendsen-Wang and Wolff cluster algorithms. The lattice sizes of
$L=10-96$ are analysed by a finite-size-scaling technique. The
site dilution concentration $p=0.85$ was chosen to minimize the
correction-to-scaling effects. We calculate numerical values of
the dynamical critical exponents for the integrated and
exponential autocorrelation times for energy and magnetization. As
expected, cluster algorithms are characterized by lower values of
dynamical critical exponent than the local one: also in the case
of dilution critical slowing down is more pronounced for the
Metropolis algorithm. However, the striking feature of our
estimates is that they suggest that dilution leads to decrease of
the dynamical critical exponent for the cluster algorithms. This
phenomenon is quite opposite to the local dynamics, where dilution
enhances critical slowing down.
\end{abstract}

\end{frontmatter}
\section{Introduction}
\label{I}

Three-dimensional random-site Ising model (random Ising model,
RIM) serves as a paradigm to describe influence of quenched
dilution on systems which in a pure, undiluted, state exhibit a
second order phase transition with scalar order parameter. The
most prominent experimental example is given by mixed crystals
${\rm Fe_pZn_{1-p}F_2}$, ${\rm Mn_pZn_{1-p}F_2}$ \cite{Belanger00}
however other physical realisations of the RIM are also possible
\cite{Folk03}. RIM offers a unique possibility to test in theory
and in simulations a change of the asymptotic critical exponents
caused by structural randomness. Indeed, due to the Harris
criterion, the new universality class does not arise in the
diluted system if the heat capacity of the corresponding pure
system does not diverge (i.e. if the critical exponent $\alpha<0$)
\cite{Harris74}. Therefore, dilution does not change universality
class of $O(m)$-symmetrical 3d systems with vector order
parameter, as easy-plane or Heisenberg magnets. Such a
universality test has been a challenge for numerous researchers in
the on-going study since late 70-ies \cite{Belanger00,Folk03}.
Summarized briefly, the main outcome of this research is that a
new universality class has been found both in theory via
renormalization group (RG) approach as well as experimentally and
by Monte Carlo (MC) simulations \cite{note1}.

However, the primary issue of the RIM studies performed so far
concerned static critical behaviour and its numerical
characteristics. Less is known about RIM critical dynamics. This
paper aims to offer extensive MC simulations of the RIM dynamical
critical properties. Moreover, as far as local and cluster MC
algorithms correspond to different forms of dynamics, a separate
task of our study is to compare numerical characteristics of
Metropolis (local) and Swendsen-Wang and Wolff (cluster) dynamics
for RIM. As to our knowledge the last question has never been
addressed so far.

The paper is organized as follows. In the next section we give a
brief summary of available theoretical and experimental data, in
section \ref{III} we introduce the model and a set of observables
we are interested in. For the sake of completeness we briefly
describe the MC algorithms in this section as well, simulation
details are summarized in the section \ref{IV}. We display and
discuss the results in sections \ref{V}--\ref{VII}.

\section{Review}
\label{II}

In the pure (undiluted) 3d Ising model the critical slowing down,
i.e. an increase of the relaxation time $\tau$ as the critical
point $T_c$ is approached, is governed by the universal dynamical
critical exponent $z$:
\begin{equation} \label{1}
\tau \sim |T-T_c|^{-\nu z},
\end{equation}
with the correlation length critical exponent $\nu$. For isotropic
systems, the dynamical exponent is related to the pair correlation
function critical exponent $\eta$ via $z=2+c\eta$
\cite{Hohenberg77}, where $c$ is a ($d$-dependent) constant. The
numerical value of the exponent $\eta$ being small, the value of
the dynamical exponent $z$ for the 3d Ising model slightly differs
from 2. Typical numbers are $z=2.1(1)$ (experiment, ${\rm FeF_2}$,
\cite{Belanger88}), $z=2.032(4)$ (MC, \cite{Grassberger95}),
$z=2.017$ \cite{Prudnikov97}, $z=2.012$ \cite{Blavatska05} (RG
theories).

Below we briefly review experimental, theoretical and MC studies
performed so far to analyse how the relation (\ref{1}) holds for
the RIM and, in particular, to answer the question how the
dynamical critical exponent $z$ is influenced by dilution.
Numerical values of $z$ that follow from our review are collected
in Table \ref{tab1}.

{\em Experiments.} There are only three independent studies of the
RIM critical dynamics we are aware of. All three concern an
antiferromagnetic uniaxial crystal ${\rm FeF_2}$ diluted by its
non-magnetic isomorph ${\rm ZnF_2}$. The resulting substance ${\rm
Fe_pZn_{1-p}F_2}$ was analysed by two different techniques:
M\"ossbauer spectroscopy (studying the dynamical line broadening
of the M\"ossbauer spectra) \cite{Barrett86,Rosov92} and by the
spin-echo neutron scattering (analyzing the time dependent spin
correlation function) \cite{Belanger88}. The earlier studies
\cite{Barrett86,Belanger88} lead to conclusion that dilution
causes a decrease of $z$: it was found to vary from 2.1 to 1.5
when magnetic atom concentration $p$ varied from 1 to 0.46
\cite{Barrett86}, whereas Ref. \cite{Belanger88} reports
$z=1.7(2)$ for $\rm{Fe_{0.46}Zn_{0.54}F_2}$ and explains it in the
frames of conventional van Hove theory: $z=2-\eta$. However, a
later experiment \cite{Rosov92} brings about the value
$z=2.18(10)$ for $\rm{Fe_{0.9}Zn_{0.1}F_2}$ being in a good
agreement with available RG results, as we will see below.

\begin{table}[htb]
\caption{The dynamical critical exponent of RIM $z$ as defined in
experiments, theory and MC simulations. The methods are given in
the following notations, experiments: MS, M\"ossbauer
spectroscopy; SENS, spin-echo neutron scattering; theory:
$\varepsilon^{1/2}$, first non-trivial order
$\varepsilon$-expansion; MRG, massive RG at $d=3$; $\overline{\rm
MS}$, minimal subtraction RG at $d=3$; Metropolis MC simulations:
FSS, finite-size-scaling; DRG, dynamical RG; OE, out-of-equlibrium
short-time dynamics.  See the text for a whole description.
\label{tab1}}
\begin{center}
{\small
\begin{tabular}{llll}
\hline
    Reference & Method & Peculiarities & $z$\\
\hline Barrett et al., 1986 \cite{Barrett86} & MS & ${\rm
Fe_pZn_{1-9}F_2}$, &  \\
  &  &  $0.46<p<1$ & $1.5<z<2.1$ \\
 Belanger et al., 1988 \cite{Belanger88} & SENS & ${\rm
Fe_{0.46}Zn_{0.54}F_2}$& 1.7(2) \\
 Rosov et al., 1992 \cite{Rosov92} & MS & ${\rm
Fe_{0.9}Zn_{0.1}F_2}$ & 2.18(10) \\
 Grinstein et al., 1977
\cite{Grinstein77} & $\varepsilon^{1/2}$ & & 2.336 \\
 Prudnikov et al., 1992 \cite{Prudnikov92} & MRG & 2 loops & 2.237 \\
 Janssen et al., 1995 \cite{Janssen95} & $\overline{\rm MS}$ & 2-3 loops & 2.18 \\
 Prudnikov et al., 1998 \cite{Prudnikov98} & MRG & 3 loops & 2.165 \\
 Blavats'ka et al., 2005 \cite{Blavatska05} & $\overline{\rm MS}$ & 2 loops & 2.172 \\
 Prudnikov, 1992 \cite{Prudnikov92a} & DRG & $L=48$, $z_{M}$ & $z(p=0.95)=2.19(7)$ \\
  &  &  & $z(p=0.8)=2.20(8)$  \\
 Heuer, 1993 \cite{Heuer93} & FSS & $L=60$, $z_{|M|, {\rm int}}$ & 2.4(1) \\
  Parisi et al., 1999 \cite{Parisi99} & OE & $L=100$, $z_{\chi}$ &
  2.62(7) \\
  Schehr et al., 2005 \cite{Schehr05} & OE & $L=100$, $z_{M}$ &
  2.6(1) \\
 \hline
\end{tabular}
}
\end{center}
\end{table}

{\em Theory.} Theoretical analysis of the RIM critical dynamics is
due to the RG approach. The majority of work was devoted to
analysis of the pure relaxational dynamics with non-conserved
order parameter described by the Langevin equation of motion,
model A dynamics in the classification of Ref. \cite{Hohenberg77}.
Being realized via MC simulations, it corresponds to the single
spin Metropolis dynamics \cite{Metropolis53}. Change of the RIM
equations of motion by coupling to the diffusive dynamics of a
conserved scalar density (energy density) does not change the
asymptotic critical behavior \cite{Lawrie84}: model C and model A
belong to the same dynamical universality class for the RIM.
However, due to crossover effects, the effective critical
behaviour essentially differs for these two forms of dynamics
\cite{Dudka05}.

The pioneering work \cite{Grinstein77} analysed by
$\varepsilon=4-d$ expansion the critical dynamics of the
$m$-vector model with quenched random impurities and non-conserved
order parameter. For the RIM case, $m=1$, new dynamical
universality class was found with the two loop value of the
exponent $z=2+\sqrt{6\varepsilon/53}$. Being analysed at $d=3$
naively by a simple substitution $\varepsilon=1$ this yields
$z=2.336$ and essentially differs from $z$ of the pure 3d Ising
model quoted at the beginning of this section. However, the RG
expansions are known to be asymptotic at best and resummation is
needed to get reliable data on their basis \cite{Zinn96}. Further
results were obtained by the massive RG approach directly at $d=3$
with subsequent resummation of resulting expansions: the two-loop
value of the exponent reads $z=2.237$ \cite{Prudnikov92}, the
three-loop one is  $z=2.165$ \cite{Prudnikov98}. Currently, due to
essential technical difficulties, dynamical RG functions of RIM
are known only within two-loop accuracy in the minimal subtraction
RG scheme  and within three loops in the massive RG approach at
$d=3$. In the minimal subtraction renormalization, the static RG
functions were taken in three loops and combined with the two-loop
expansions for the dynamic ones. These gave the following
$\varepsilon$-expansion for the dynamical exponent:
$z=2+0.336\sqrt{\varepsilon}(1-0.932\sqrt{\varepsilon})$ with the
naive estimate $z=2.023$ \cite{Oerding95}. Being improved by the
resummation of the static RG functions, the estimate reads
$z=2.18$ \cite{Janssen95}. The last value is close to the other
estimate, obtained from the resummation of the two-loop minimal
subtraction RG functions directly at $d=3$: $z=2.172$
\cite{Blavatska05}.

Theoretical studies mentioned above concerned dynamic criticality
associated with equilibrium fluctuations. Recently, it was shown
\cite{Janssen89} that non-equilibrium relaxation at short times
possess scaling features as well. In particular, if a system is
suddenly quenched from high temperatures to the critical one and
then released to the dynamic evolution of model A, the evolution
at short times is governed by scaling laws. Both dynamical and
static critical exponents can be extracted from the scaling.
Short-time dynamics of the RIM at non-equilibrium critical
relaxation was analysed if Refs.
\cite{Kissner92,Oerding95,Schehr05,Schehr05a}.

Let us mention  a related problem, where an influence of quenched
disorder on RIM critical dynamics was examined theoretically.
These are studies of an effect of extended impurities on RIM
critical dynamics \cite{Lawrie84,extended,Dudka05}. As far as the
presence of extended (long-range correlated) impurities changes
the static universality class of RIM, the dynamical critical
behaviour is found to differ from those of the RIM with point-like
uncorrelated disorder.

{\em MC simulations.} Essential progress in MC simulations of
static critical phenomena is due to the application of cluster
algorithms \cite{Swendsen87,Wolff89}. In particular, they allowed
to obtain precise values of the RIM static critical exponents
\cite{Folk03}. It is to be emphasized here, that whereas the
cluster algorithms were specially designed to lead to the same
static critical behaviour as the single-spin Metropolis algorithm
\cite{Metropolis53}, it is not the case for dynamics. It is the
Metropolis algorithm (due presumably to its locality), which leads
to the same value of dynamic critical exponent as the one observed
for RIM experimentally in Refs.
\cite{Belanger88,Barrett86,Rosov92} and analysed theoretically in
Refs.
\cite{Grinstein77,Prudnikov92,Prudnikov98,Oerding95,Janssen95,Blavatska05}.
In cluster algorithms, as it follows already from their name, the
whole clusters of spins are flipped, which gives origin to the
non-local dynamics. In its turn, the last is characterized by its
own dynamical critical exponents
\cite{Li89,Coddington92,Ossola04}. As far as the cluster
algorithms were introduced to overcome the critical slowing down,
the corresponding autocorrelation times are characterized by
weaker singularities as those of local dynamics: $z_{\rm
cluster}<z_{\rm local}$.

Let us note, that the theoretical RG calculations assume a single
dynamical critical exponent $z$ for the relaxation times of
different observables. In the MC simulations one typically finds,
that the autocorrelation time of different observables is
characterized by different (effective) exponents, which are
expected to coincide in the asymptotics. Therefore, when we give
the MC values of $z$ in Table~\ref{tab1} we specify also the
physical observable for which it has been measured. Already the
first MC study of the RIM single-spin critical dynamics revealed
dynamical scaling behaviour with a concentration-dependent
critical exponent $z$ \cite{Heuer93}. For small dilution the
following numbers were reported: $z(p=0.95)=2.15(1)$;
$z(p=0.9)=2.23(1)$; $z(p=0.8)=2.39(1)$. The concentration
dependence of $z$ was explained by crossover. In the region of
concentrations $p\simeq0.8$ the slope of the crossover function
was found to change its sign, therefore the correction-to-scaling
terms were minimal. This allowed to arrive to the conclusion about
an asymptotic value of the dynamic critical exponent $z=2.4(1)$
\cite{Heuer93}.  Independently, Metropolis dynamics of the RIM was
analysed in Ref. \cite{Prudnikov92a} by a combination of MC and
dynamical RG \cite{Jan83} techniques. Again the
concentration-dependent exponents were found with a different
conclusion, however: a hypothesis of RIM step-like universality
was proposed. According to the hypothesis, the asymptotic critical
exponents remain unchanged only within certain concentration
region. For a small dilution the values of $z$ practically did not
differ: $z(p=0.95)=2.19(7)$; $z(p=0.8)=2.20(8)$
\cite{Prudnikov92a} and are compatible with those obtained by a
finite-size-scaling technique in Ref. \cite{Heuer93}.

Other estimates come from MC simulations of the out-of-equilibrium
RIM dynamics. Here, taking into account correction-to-scaling,
value $z=2.62(7)$ was extracted from the time dependence of the
out-of-equilibrium susceptibility $\chi(t)$ (with the leading
dynamical correction-to-scaling exponent $\omega=0.50(13)$)
\cite{Parisi99}. This value was further supported by the
out-of-equilibrium simulations of Ref. \cite{Schehr05}, where the
value $z=2.6(1)$ was extracted from the scaling of the spin-spin
autocorrelation function.

As it was noted above, the MC cluster algorithms provide different
type of dynamics and therefore their scaling exponents can not be
compared straightaway with those of single-spin local dynamics
summarized in Table \ref{tab1}. Moreover, currently there is no
field theory available to predict the critical exponent value for
cluster dynamics even for the pure (undiluted) spin models. Study
of such dynamics constitutes a separate task and certain
analytical and numerical work has already been done for the pure
models \cite{Li89,Coddington92,Ossola04}. As to our knowledge, no
results for the RIM have been obtained so far. An exception is
Ref. \cite{Berche04}, where an effective (concentration dependent)
critical exponent $z$ was obtained for the Swendsen-Wang cluster
algorithm for 3d random-bond Ising model. For a small bond
dilution an estimate reads: $z(p=0.7)=0.41$ \cite{Berche04}.

\section{Observables and MC algorithms}
\label{III}

In our paper we consider the $3d$ Ising model with non-magnetic
impurities randomly distributed over the system. The Hamiltonian
of this model on the cubic lattice has the following form

\begin{equation}
{\mathcal H} = -J\sum_{\langle ij\rangle }c_ic_jS_iS_j,
\label{Ham}
\end{equation}
where $\langle ij\rangle $ denotes the summation over the nearest
neighbour sites of the lattice, $c_i=1$ if the $i$-th site is
occupied by a spin and $c_i=0$ otherwise, the Ising spins $S_i$
take on the values $+1$ or $-1$. The spins interact via an
exchange coupling $J$, which is positive. Occupied sites ($c_i=1$)
are considered to be uncorrelated, randomly distributed and
quenched in a fixed configuration. For every observable discussed
below, first the Boltzmann average with respect to the spin
subsystem is performed for the fixed disorder realisation, then
the averaging over different disorder realisations is performed.
We will use the following notations: Boltzmann average over the
spin subsystem will be denoted by angular brackets
$\langle(\dots)\rangle$ whereas the over bar $\overline{(\dots)}$
will stand for the averaging over the disorder realisations. The
number of all sites is $N=L^3$ and the number of sites carrying a
spin is $N_p$. The concentration of spins is defined therefore as
$p = N_p/N$.

\subsection{The properties of interest} \label{IIIa}

For a given disorder realisation, an average value of an
observable $\langle{\mathcal O}\rangle$ at temperature $T$ can be
computed in the canonical ensemble from its values ${\mathcal O}$
for given spin configurations:
\begin{equation}
\langle{\mathcal O}\rangle = \frac{1}{\mathcal Z}{\rm
Sp}\,{\mathcal O} e^{-\beta { \mathcal H}},
\end{equation}
where $\beta=(kT)^{-1}$,  ${\mathcal Z}$ is the partition function
\begin{equation}\label{part}
{\mathcal Z}={\rm Sp}\, e^{-\beta { \mathcal H}},
\end{equation}
and trace in (\ref{part}) is taken over the spin degrees of
freedom. In the course of the MC simulation each spin
configuration is generated with its proper Boltzmann weight
already (more detailed description of the algorithms is given in
the next section), hence the thermodynamic average is the simple
average over all generated configurations. If $N_{\rm steps}$ is
the total number of productive MC steps used for the averaging,
then
\begin{equation}
\langle {\mathcal O} \rangle = \frac{1}{N_{\rm steps}}\sum_{\rm
conf} {\mathcal O}. \label{q}
\end{equation}
Sum in (\ref{q}) spans all spin configurations in which the (spin
configuration dependent) observable ${\mathcal O}$ is measured.

In an ideal case of uncorrelated, statistically independent
configurations, the total error in defining $\langle{\mathcal
O}\rangle$ can be evaluated as
\begin{equation}
\langle \delta {\mathcal O} \rangle =\sqrt{\frac{\sum_{\rm conf}
 (\delta {\mathcal O})^2}{N_{\rm steps}(N_{\rm
steps}-1)}}, \label{deltaq}
\end{equation}
where $\delta {\mathcal O}= {\mathcal O}-\langle{\mathcal
O}\rangle$. In practice, however, the correlation between
different spin configurations exists as the result of particular
MC scheme. This correction can be characterized via the (disorder
dependent) autocorrelation function \cite{Allen87}:
\begin{equation}\label{aut}
C_{\mathcal O}(\delta t) = \frac{\langle\delta {\mathcal
O}(t_0+\delta t)\delta {\mathcal O}(t_0)\rangle} {\langle \delta
{\mathcal O}(t_0+\delta t)\rangle \langle \delta {\mathcal
O}(t_0)\rangle},
\end{equation}
where $t_0$ is some time origin.

At times large enough $C_{\mathcal O}(\delta t)$ decays
exponentially according to the Debye law
\begin{equation}\label{tauE}
C_{\mathcal O}(\delta t) = a\,e^{-\delta t/\tau_{{\mathcal O},
{\rm exp}}},
\end{equation}
where $\tau_{{\mathcal O},{\rm exp}}$ is the exponential
autocorrelation time, obtained for quantity ${\mathcal O}$ and $a$
is a constant. Time $\tau_{{\mathcal O},{\rm exp}}$ defines a time
scale at which the configurations generated in a course of the MC
simulation can be assumed as uncorrelated. Hence, in the
simulation run of length $N_{\rm steps}$ MC steps, only
$\frac{N_{\rm steps}}{2\tau_{{\mathcal O}, {\rm exp}}}$
configurations are considered to be statistically independent.

Besides the  $\tau_{{\mathcal O},{\rm exp}}$, the relaxation of an
observable ${\mathcal O}$ is characterized by the integrated
autocorrelation time $\tau_{{\mathcal O},{\rm int}}$. It is
defined via
\begin{equation}\label{tauE1}
\tau_{{\mathcal O},{\rm int}} = \frac{1}{2}+\sum_{\delta
t=1}^\infty C_{\mathcal O}(\delta t)).
\end{equation}
In practice, $\tau_{{\mathcal O}, {\rm int}}$ is evaluated by
introducing a maximum cutoff in the sum (\ref{tauE1}) and it may
be shown that both autocorrelation times coincide $\tau_{{\mathcal
O},{\rm int}}=\tau_{{\mathcal O},{\rm exp}}$ only in the limit
when this cutoff goes to $\delta t\rightarrow \infty$, otherwise
$\tau_{{\mathcal O},{\rm int}} < \tau_{{\mathcal O},{\rm exp}}$
\cite{Janke02}.

Let us specify now the observables we will be interested in during
MC simulations. In this study we concentrate on the internal
energy ${\mathcal E}$, the magnetization ${\mathcal M}$ and
absolute value of the magnetization $|{\mathcal M}|$ per site,
defined as
\begin{equation}
{\mathcal E} = - J\frac{1}{N_p}\sum_{\langle ij\rangle
}c_ic_jS_iS_j, \label{Ene}
\end{equation}
\begin{equation}
{\mathcal M} = \frac{1}{N_p}\sum_{i}c_iS_i, \label{Mag}
\end{equation}
\begin{equation}
|{\mathcal M}| = \frac{1}{N_p}|\sum_{i}c_iS_i|. \label{Mag_abs}
\end{equation}
From these observables (${\mathcal O}$) we compute the following
expectation values ($O$):
\begin{equation}
E=\overline{\langle{\mathcal E}\rangle}, \hspace{2em} M=
\overline{\langle{\mathcal M}\rangle}, \hspace{2em} |M|=
\overline{\langle|{\mathcal M}|\rangle}.
\end{equation}

The divergency of the autocorrelation time for any of these
quantities as the $T_c$ is approached, Eq. (\ref{1}), for a finite
system of size $L$ is reflected in a power law scaling of $\tau$
with $L$. Taken that both exponential and integrated
autocorrelation times are defined during simulation, their scaling
is governed by corresponding exponents:
\begin{eqnarray}\label{fff1}
\tau_{ O,{\rm exp}} \sim L^{z_{ O,{\rm exp}}},\\ \label{fff2}
\tau_{ O,{\rm int}} \sim L^{z_{ O,{\rm int}}},
\end{eqnarray}
with $O$ being any of the expectation values $E$, $M$, $|M|$
computed in simulations.

Note, that the theoretical RG calculations assume a unique
dynamical critical exponent $z$ for the relaxation times of all
observable quantities (cf. theoretical estimates for $z$ in
Table~\ref{tab1}). This might not hold for the scaling of the
experimentally defined autocorrelation times (\ref{fff1}),
(\ref{fff2}). For the pure system, it is believed that the scaling
of autocorrelation times for the energy-like observables is
described with the same dynamical critical exponent
\cite{Salas97}, the last might differ from that for the
susceptibility-like observables \cite{Ossola04}.

As it was noted already above, local and cluster MC algorithms
give rise to different forms of critical dynamics. Therefore, they
are described by different scaling exponents (\ref{fff1}),
(\ref{fff2}). Indeed, the large clusters of equally oriented spins
are started to be formed in the vicinity of a critical point.
Therefore, most of the spin update attempts made by any of local
algorithms (e.g. by the Metropolis one) are wasted, and, as
result, the generated configurations are highly correlated. Both
the correlation time $\tau$ and the critical index $z$ are large
and system moves in a configurational space inefficiently. This
poses the severe difficulty for the calculation of the static
critical exponents, but, in fact, reflects the real dynamics in
the system near the critical point.

The non-local, cluster algorithms introduced by Swendsen-Wang and
Wolff consider cluster pseudo-dynamics of the system by attempting
to flip the whole cluster(s) at once. The primary goal is to
reduce the effective autocorrelation time and, therefore, to
improve the statistical sampling of generated configurations. At
the same time, these algorithms bring the system into different
dynamical class and an analysis of the different dynamics governed
by different MC algorithms is the main goal of this study. Below,
for the sake of completeness, we briefly describe main steps of
the MC algorithms under consideration.

\subsection{MC algorithms} \label{IIIb}

The Metropolis algorithm is the simplest and historically the
first MC algorithm \cite{Metropolis53}. It utilises the
preferential sampling of the configurational space, where each
spin configuration is generated with appropria\-te Boltzmann
weight. The technical difficulty of generating all the new
configurations independently is overcame by using instead the
recipe how to produce each new configuration from the previous
one. Spins are randomly selected and flipped with a probability
$P=min(1,e^{-\beta\Delta H})$ where $\Delta H$ is the energy
difference between the old and the new configurations. A Markovian
chain of configurations is thus produced and defines the
pseudodynamics of the system.

Locality of this algorithm is a serious drawback near the critical
point, where large correlated clusters of spins are emerging. As
the result, the vast amount of single spin flips are rejected,
therefore the configurations generated are highly correlated and
the system moves in a phase space inefficiently.

Cluster algorithms were designed to overcome these difficulties.
They are based on the identification of clusters of sites using a
bond percolation process connected to the spin configuration of
the magnetic system. All spins of the clusters are then
independently flipped.

In the case of the Ising model (or more generally the Potts
model), the percolation process involved is obtained through the
mapping onto the random graph model, first addressed by Fortuin
and Kasteleyn~\cite{Fortuin}. In the Swendsen-Wang algorithm
\cite{Swendsen87}, a cluster update sweep consists of the
following steps: depending on the nearest neighbour exchange
interactions and site occupations, assign value to a bond between
sites $i$ and $j$ with probability $P_{ij}=1-e^{-2\beta Jc_ic_j}$,
then identify clusters of spins connected by active bonds, and
eventually assign a random value to all the spins in a given
cluster.  The spin system at criticality is mapped into a bond
percolation problem at the percolation threshold. It results in
the producing of clusters of arbitrary large sizes, therefore the
Swendsen-Wang algorithm samples the configurations in a critical
region much more efficiently. This is reflected in its rather
small  value of the dynamical critical exponent $z$. Some
disadvantage is an extra computer time required to split-up the
system into clusters.

Wolff introduced a single cluster algorithm~\cite{Wolff89} which
otherwise is much similar to the Swendsen-Wang one. A spin is
randomly chosen, then the cluster connected with this spin is
constructed and all the spins in the cluster are updated. Note
that in this scheme, the flip of one cluster updates only the
spins belonging to this cluster and therefore produces only a
partial update of the system. To match the same time scale as in
the Metropolis and Swendsen-Wang algorithms, the time scale of the
Wolff algorithm should be corrected by a factor $c= l_{\rm
cluster}/L^3$, where $l_{\rm cluster}$ is the average size of
flipped clusters.

\section{Simulation details}
\label{IV}

In the rest of the paper we study RIM critical dynamics governed
by the three different MC algorithms described in the previous
section. The MC simulations are performed for a range of system
sizes up to $L=96$ with periodic boundary conditions. The
concentration of magnetic sites was fixed at $p=0.85$. This choice
is based on the previous findings that the correction-to-scaling
terms are minimal at concentrations $p\sim 0.8$
\cite{Heuer93,Ballesteros98}. Therefore, we do not account for
these terms. The simulations are performed at the temperature that
corresponds to the critical temperature of the infinite system and
was taken equal to $\beta_c J=0.2661922(83)$ according to our
previous findings \cite{Ivaneyko05}.

\begin{table}[htb]
\caption{CPU time (in seconds) used for performing 1000 MC steps
and 10 disorder realisations.} \label{tab2}
\begin{center}
\begin{tabular}{lllllllll}
\hline
 $L$ & 10 & 12 & 16 & 24 & 32 & 48 & 64 & 96 \\
 \hline
 Metropolis &
3.33&5.81&13.77&50.17&119.26&403.69&957.72&3207.54\\
 Sw.-Wang &
3.60&6.18&14.60&55.10&130.78&442.64&1092.76&3682.48\\
 Wolff&1.25&2.05&4.53&18.11&43.39&139.81&298.33&970.05\\
 \hline
\end{tabular}
\end{center}
\end{table}

Typically, we averaged over $N_{\rm dis}=10^3$ disorder
realisations for each lattice  size. All runs were started from a
random configuration of empty and occupied spin sites. At first,
we run $~250\tau_{E,{\rm int}}$ MC sweeps for thermal
equilibration and then the production run of $10^4\tau_{E,{\rm
int}}$ MC sweeps was conducted. For all cases that was quite
sufficient for the accurate description of the long-time behaviour
of the autocorrelation functions.

As a basis for the random number generator we take minimal random
number generator $Ran1$ of Park and Miller with Bays-Durham
shuffle and added safeguards, described in Ref. \cite{Flannery}.

We used the workstations cluster of the ICMP based on Athlon MP
2200+ processors. The typical simulation time per 1000 MC sweeps
and 10 disorder realisations with data records are shown in the
Table.~\ref{tab2}.

\section{Autocorrelation times}
\label{V}

\begin{figure}[ht]
\epsfxsize=8cm \centerline{\epsffile{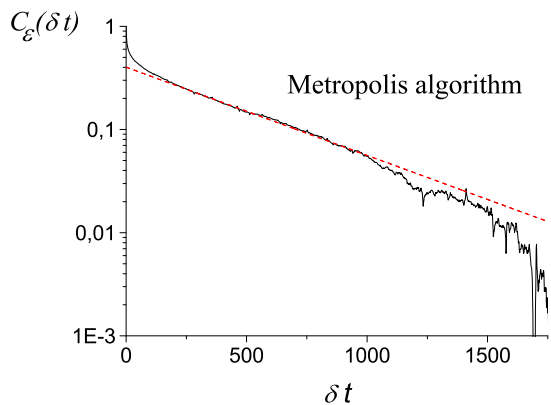}} \caption{The
log-linear plot for the energy-energy autocorrelation function
$C_{{\mathcal E}}(\delta t)$, the Metropolis algorithm, $L=64$.
 Bold line: measured value. Dashed
line: fit to the exponential decay (\ref{tauE}) with the
autocorrelation time $\tau_{{\mathcal E},{\rm exp}}=507.8$.
\label{fig2}}
\end{figure}

\begin{figure}[ht]
\epsfxsize=8cm \centerline{\epsffile{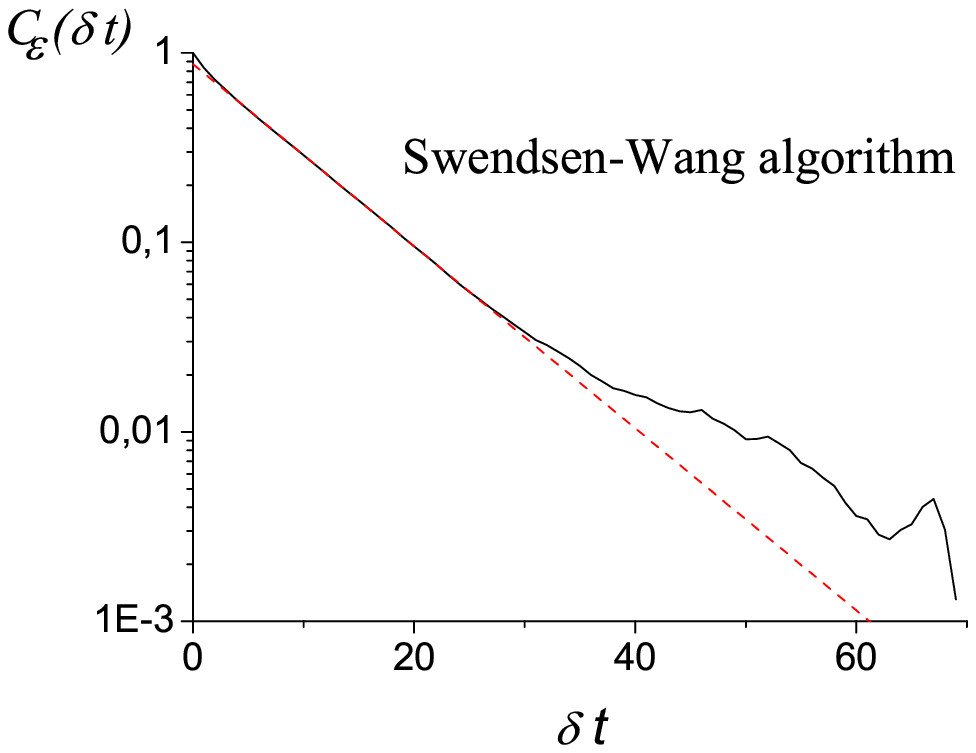}} \caption{The
log-linear plot for the energy-energy autocorrelation function
$C_{\mathcal E}(\delta t)$, the Swendsen-Wang algorithm, $L=64$.
Bold line: measured value. Dashed line: fit to the exponential
decay (\ref{tauE}) with the autocorrelation time $\tau_{{\mathcal
E},{\rm exp}}=9.04$. \label{fig3}}
\end{figure}

\begin{figure}[ht]\epsfxsize=8cm
\centerline{\epsffile{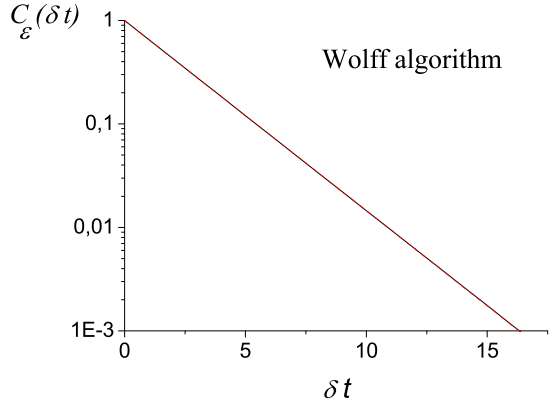}} \caption{ The log-linear plot for
the the energy-energy autocorrelation function $C_{\mathcal
E}(\delta t)$, the Wolff algorithm, $L=64$.  The measured value
and the fit to the exponential decay (\ref{tauE}) with the
autocorrelation time $\tau_{{\mathcal E},{\rm exp}}=2.63$ are
indistinguishable within the scale chosen. \label{fig4}}
\end{figure}

When performing the MC simulations on a system of $N$ spins, the
time unit can be related to the number of MC sweeps (MCS), where
during one sweep on average $N$ spins are updated. This convention
is also used in our study. A quantitative analysis of the
autocorrelation times involves an evaluation of the
autocorrelation function for various properties of the system
\cite{Allen87}. As was mentioned, in this study we concentrate on
the autocorrelations of the energy $\mathcal{E}$, the
magnetization $\mathcal{M}$ and the absolute value of
magnetization $|\mathcal{M}|$. Defined in (\ref{aut}), the
autocorrelation function can be rewritten as:
\begin{eqnarray}
C_{{\mathcal O}}(\delta t) &=& \frac{\langle\delta {\mathcal
O}(t_0+\delta t)\delta {\mathcal O}(t_0)\rangle} {\langle \delta
{\mathcal O}(t_0+\delta t) \rangle \langle \delta {\mathcal
O}(t_0)\rangle}=\nonumber\\ &=&\frac{\langle {\mathcal
O}(t_0+\delta t){\mathcal O}(t_0)\rangle-\langle {\mathcal
O}(t_0+\delta t)\rangle \langle {\mathcal O}(t_0)\rangle}
{\sqrt{(\langle {\mathcal O}(t_0+\delta t)^2\rangle-\langle
{\mathcal O}(t_0+\delta t)\rangle^2)(\langle {\mathcal
O}(t_0)^2\rangle-\langle {\mathcal O}(t_0)\rangle^2)}} .
\end{eqnarray}
In the thermodynamic limit $\langle {\mathcal O}(t_0+\delta
t)\rangle$ is equal to $\langle {\mathcal O}(t_0)\rangle$, and one
arrives to the more usual expression
\begin{equation}
C_{{\mathcal O}}(\delta t) = \frac{\langle {\mathcal
O}(t_0){\mathcal O}(t_0+\delta t)\rangle - \langle {\mathcal
O}(t_0)\rangle \langle {\mathcal O}(t_0)\rangle } {\langle
{\mathcal O}(t_0){\mathcal O}(t_0)\rangle -\langle {\mathcal
O}(t_0)\rangle \langle {\mathcal O}(t_0)\rangle },
\end{equation}
where ${\mathcal O}(t)$ is the instant value for the property of
interest at certain time $t$, ($t_0$ is some time origin, $ \delta
t$ is the time elapsed since the time origin $t_0$). The averaging
over a large number of time origins $t_0$ is needed to smoothen up
the $C_{{\mathcal O}}(\delta t)$ at large $\delta t$. As an
example, a typical behaviour of  the energy-energy autocorrelation
functions for given disorder realisations is shown in
Figs.~\ref{fig2}-\ref{fig4}.

\begin{table}[htb]
\caption{The average cluster size $l_{\rm cluster}$ and factor $c$
for Wolff MC algorithm} \label{tab3}
\begin{center}
\begin{tabular}{lllllllll}
\hline
 $L$ & 10 & 12 & 16 & 24 & 32 & 48 & 64 & 96 \\
 \hline
$l_{\rm cluster}$& 149&216&385&856&1526&3393&6241&15034\\
 $c$ & 0.175 & 0.147 &0.111&0.073&0.055& 0.036 &0.028& 0.020\\
 \hline
\end{tabular}
\end{center}
\end{table}

As was already mentioned above, the time scale of the Wolff
algorithm should be accounted for the average cluster size to be
compared correctly with the dynamics of other algorithms. To this
end, for each lattice size and for a given disorder realisation we
calculated the size of flipped cluster, skipping the first
$~250\tau_{E, {\rm int}}$ MC steps for thermal equilibration. Then
we performed configurational averaging of the updated cluster
size. The scaling factor $c$ was calculated as $c = l_{\rm
cluster}/N_p$, where $l_{\rm cluster}$ is the cluster size
averaged over different disorder realisations. In
Table~\ref{tab3}, the average cluster size and the scaling factor
$c$ are presented. One could explain the behaviour of the scaling
factor $c$ by the following considerations. For the smaller system
sizes, the simulation temperature (which is equal to
$T_c^{\infty}$) is much lower than the effective critical
temperatures $T_c^{L}$, the system is at $T<T_c^{L}$ so that the
typical cluster occupy larger part of the system. With the
increase of the system size, the simulation temperature is getting
closer to the $T_c^{L}$ and the average clusters being flipped are
decreasing in size.

To calculate the  integrated autocorrelation time $\tau_{{\mathcal
O}, {\rm int}}$ the expression (\ref{tauE1}) is used. One should
note that the error for the autocorrelation function is always
larger at long times, where the data are averaged over less
intervals. The compromised accuracy for the integrated
autocorrelation time can be achieved by using certain time cutoff
$\delta t_{max}$, typically of the order of $\delta t_{max}\geq
6\tau_{int}$ \cite{Janke02}:
\begin{equation}\label{tauE2}
\tau_{{\mathcal O}, {\rm int}}(\delta t_{max}) =
\frac{1}{2}+\sum_{\delta t=1}^{\delta t_{max}}C_{\mathcal
O}(\delta t)),
\end{equation}
The trailing part  of the autocorrelation function for $\delta t >
\delta t_{max}$ can be approximated by an exponential function. As
the result, the final expression reads
\begin{eqnarray}\label{tau}
\tau_{{\mathcal O}, {\rm int}} = \frac{1}{2}+\sum_{\delta
t=1}^{\delta t_{max}} C_{\mathcal O}(\delta t))+a\sum_{\delta
t=\delta t_{max}+1}^{\infty}e^{-\delta t/t_{{\mathcal O}, {\rm
exp}}}=\nonumber\\ =\tau_{{\mathcal O}, {\rm int}}(\delta
t_{max})+a\frac{e^{-1/\tau_{{\mathcal O}, {\rm
exp}}}}{1-e^{-1/\tau_{{\mathcal O}, {\rm exp}}}} e^{-\delta
t_{max}/\tau_{{\mathcal O}, {\rm exp}}}.
\end{eqnarray}

In this study, we employed the following scheme. For each disorder
realisation the autocorrelation function $C_{\mathcal O}(\delta
t)$ has been evaluated first. To calculate first term in
(\ref{tau}) we use (\ref{tauE2}) with condition of the cutoff:
$\delta t_{max}\geq 6\tau_{{\mathcal O}, {\rm int}}$. To evaluate
the second term, one has to estimate $\tau_{{\mathcal O}, {\rm
exp}}$. Whereas the pure systems display asymptotic behaviour
dominated by a single relaxation time, the distinct feature of the
autocorrelation function of disordered systems is that there is a
whole spectrum of autocorrelation times in the crossover region
and it is reflected in the curvature of the autocorrelation
functions in the log-log plot. This feature has been noted already
in Ref. \cite{Heuer93}. Therefore, we plot the $C_{\mathcal O
}(\delta t)$ in log-log scale and estimated  $\tau_{{\mathcal
O},{\rm exp}}$ and $a$ from the straight line region in a window
$\tau_{{\mathcal O}, {\rm int}}(\delta t_{max})$ to
$3\tau_{{\mathcal O },{\rm int}}(\delta t_{max})$. The averaged
over disorder realizations exponential autocorrelation time
$\tau_{O, {\rm exp}}$ is given in Table~\ref{tab4} .

\begin{table}[htb]
\caption{Exponential autocorrelation times $\tau_{ O, {\rm exp}}$
of RIM at the critical temperature of infinite system for
different lattice sizes $L$ measured in MC sweeps for Metropolis,
Swendsen-Wang and  Wolff MC algorithms. \label{tab4}}
\begin{center}
 \tabcolsep0.3mm
{\small
\begin{tabular}{llllllll}
 \hline \multicolumn{1}{l}{$L$\hspace*{5mm}}
&\multicolumn{3}{l}{Metropolis}&
\multicolumn{2}{l}{Swendsen-Wang}&\multicolumn{2}{l}{Wolff} \\
\hline &$\tau_{E,{\rm exp}}$&$\tau_{|M|,{\rm exp}}$&$\tau_{M,{\rm
exp}}$&$\tau_{E,{\rm exp}}$& $\tau_{|M|,{\rm exp}}$&$\tau_{E,{\rm
exp}}$&$\tau_{|M|,{\rm exp}}$\\
 \hline
10&8.94(63)            &9.34(88) &6.80(1.80)$\cdot 10
$&3.76(19)&3.72(20)&1.53(17)&1.23(10)\\
 12&1.33(11)$\cdot 10$ &1.40(16)$\cdot 10$&1.02(28)$\cdot 10^2$ &4.16(23)&4.12(21)&1.64(17)&1.30(9)\\
 16&2.49(24)$\cdot 10$ &2.69(44)$\cdot 10$&1.98(63)$\cdot 10^2$ &4.82(27)&4.80(36)&1.80(21)&1.41(10)\\
 24&5.99(37)$\cdot 10$ &6.51(95)$\cdot 10$&5.10(19)$\cdot 10^2$ &5.87(38)&5.92(37)&2.06(24)&1.54(10)\\
 32&1.12(08)$\cdot 10^2$ &1.24(28)$\cdot 10^2$&1.02(57)$\cdot 10^3$ &6.78(52)&6.71(42)&2.20(28)&1.65(11)\\
 48&2.71(25)$\cdot 10^2$  &2.86(40)$\cdot 10^2$&2.40(1.30)$\cdot 10^3$ &8.15(69)&8.02(52)&2.51(30)&1.75(10)\\
 64&4.65(48)$\cdot 10^2$ &6.31(97)$\cdot 10^2$ &4.78(2.25)$\cdot 10^3$ &9.19(85)&9.03(76)&2.35(21)&1.88(11)\\
 96&1.22(14)$\cdot 10^2$ &1.44(26)$\cdot 10^3$ &1.20(0.52)$\cdot 10^4$ &10.7(9)&10.5(9)&2.79(35)&1.99(16)\\
 \hline
\end{tabular}
}
\end{center}
\end{table}

\begin{table}[htb]
\caption{Integrated autocorrelation times $\tau_{O, {\rm int}}$ of
RIM at the critical temperature of infinite system for different
lattice sizes $L$ measured in MC sweeps for Metropolis,
Swendsen-Wang and Wolff MC algorithms. \label{tab5} }
\begin{center}
\tabcolsep0.3mm
 {\small
\begin{tabular}{llllllll}
 \hline \multicolumn{1}{l}{$L$} \hspace*{5mm}
&\multicolumn{3}{l}{Metropolis}&
\multicolumn{2}{l}{Swendsen-Wang}&\multicolumn{2}{l}{Wolff} \\
\hline &$\tau_{E,{\rm int}}$&$\tau_{|M|,{\rm int}}$&$\tau_{M,{\rm
int}}$&$\tau_{E,{\rm int}}$& $\tau_{|M|,{\rm int}}$&$\tau_{E,{\rm
int}}$&$\tau_{|M|,{\rm int}}$\\
 \hline
10&6.14(30)          &9.22(49)            &6.6(1.6)$\cdot
10$&3.57(16)&3.22(17)&1.33(8)&0.95(4)\\
 12&8.47(49)          &1.36(8)$\cdot 10$   &9.8(2.5)$\cdot 10$&3.92(18)&3.50(19)&1.42(9)&0.98(4)\\
 16&1.44(10)$\cdot 10$&2.53(20)$\cdot 10$  &1.87(53)$\cdot 10^2$&4.47(23)&3.95(25)&1.55(12)&1.00(4)\\
 24&3.10(20)$\cdot 10$&6.13(48)$\cdot 10$  &4.74(1.50)$\cdot 10^2$&5.33(31)&4.68(27)&1.75(15)&1.02(4)\\
 32&5.51(40)$\cdot 10$&1.16(14)$\cdot 10^2$&9.05(3.41)$\cdot 10^2$&6.05(36)&5.14(34)&1.87(18)&1.04(4)\\
 48&1.22(12)$\cdot 10^2$&2.71(28)$\cdot 10^2$&2.17(0.82)$\cdot 10^3$&7.11(47)&6.07(43)&2.08(20)&1.04(4)\\
 64&2.15(20)$\cdot 10^2$&5.59(62)$\cdot 10^2$&3.80(1.44)$\cdot 10^3$&7.81(51)&6.71(50)&2.16(18)&1.07(3)\\
 96&4.84(51)$\cdot 10^2$&1.24(15)$\cdot 10^3$&9.57(3.45)$\cdot 10^3$&9.21(61)&7.65(59)&2.33(25)&1.10(4)\\
 \hline
\end{tabular}
}
\end{center}
\end{table}
In order to calculate the error bars for the exponential and
integrated relaxation times we use the blocking method. We divide
all set of autocorrelation times (each corresponding to a separate
replica) into $n$ blocks so that each block contains $\frac{N_{\rm
dis}}{n}$ values of the autocorrelation times. We obtain the
average value and standard error due to formulas (\ref{q}) and
(\ref{deltaq}). Then we do simple averaging over $n$ blocks. We do
not  give results for $\tau_{M,{\rm exp}}$ and $\tau_{M, {\rm
int}}$ for cluster methods, because the correlation of $M$ are
absent.

\section{Critical exponents}
\label{VI}

\begin{figure}[ht] \epsfxsize=7cm \epsffile{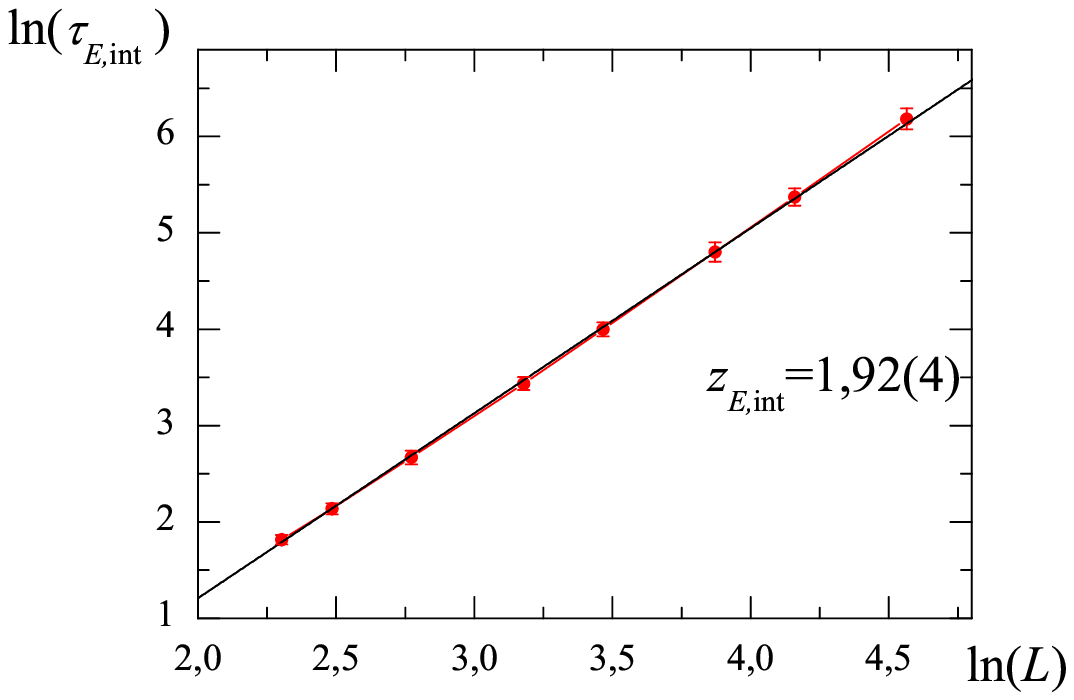}
\epsfxsize=7cm \epsffile{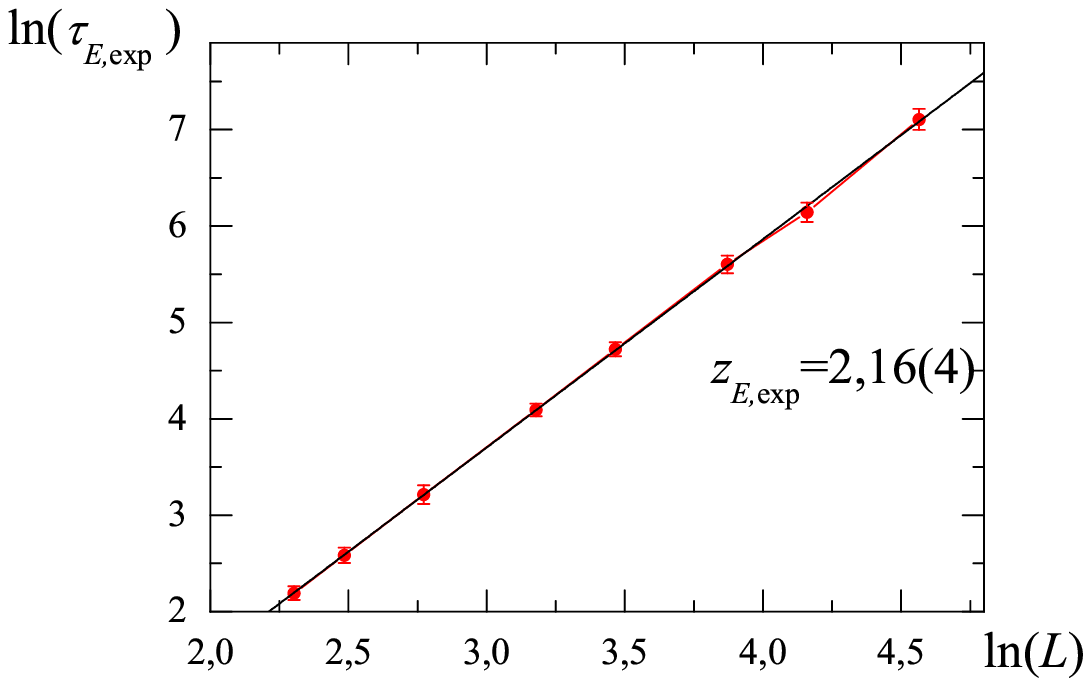}
\centerline{(a)\hspace{8cm}(b) } \epsfxsize=7cm
\epsffile{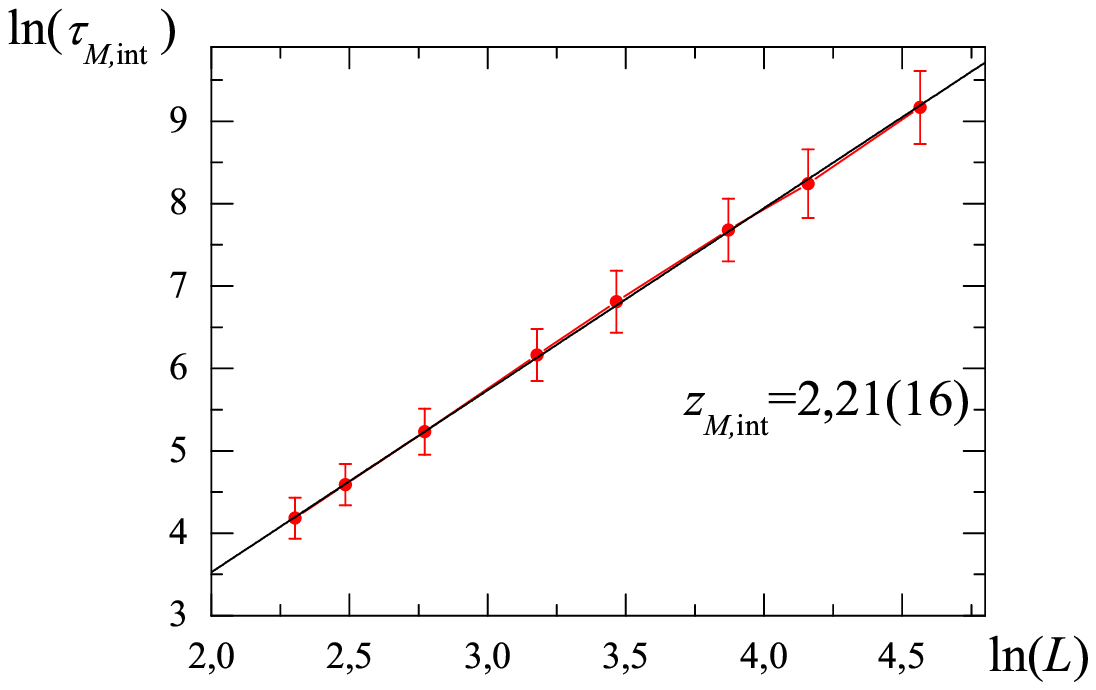} \epsfxsize=7cm \epsffile{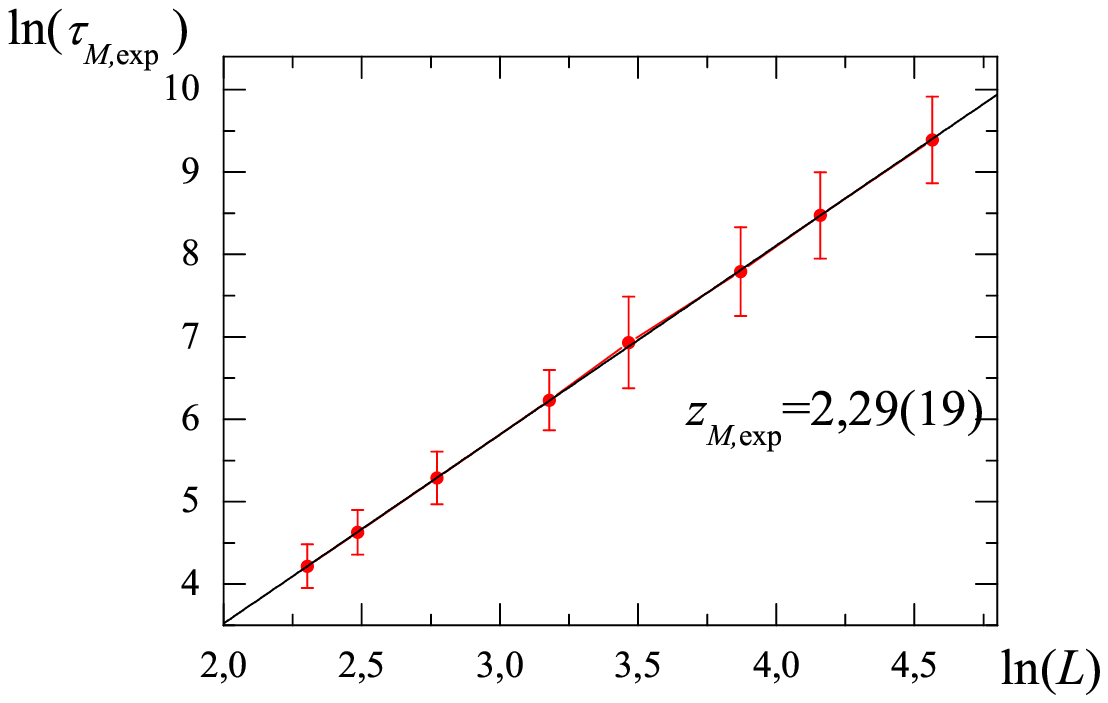}
\centerline{(c)\hspace{8cm}(d) }
\epsfxsize=7cm
\epsffile{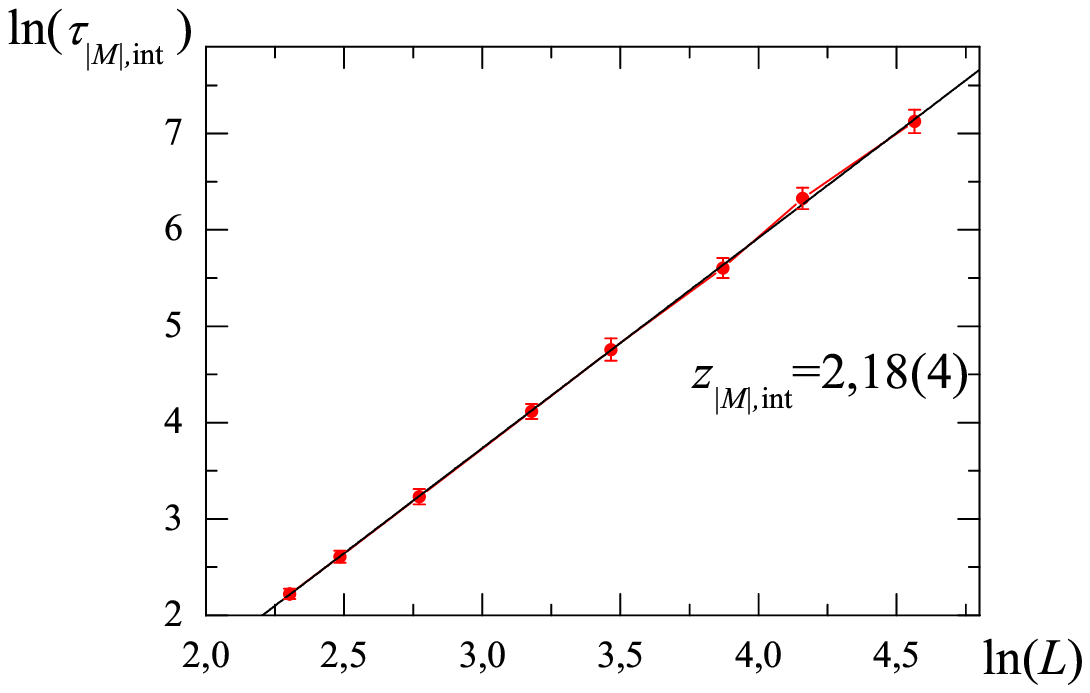} \epsfxsize=7cm \epsffile{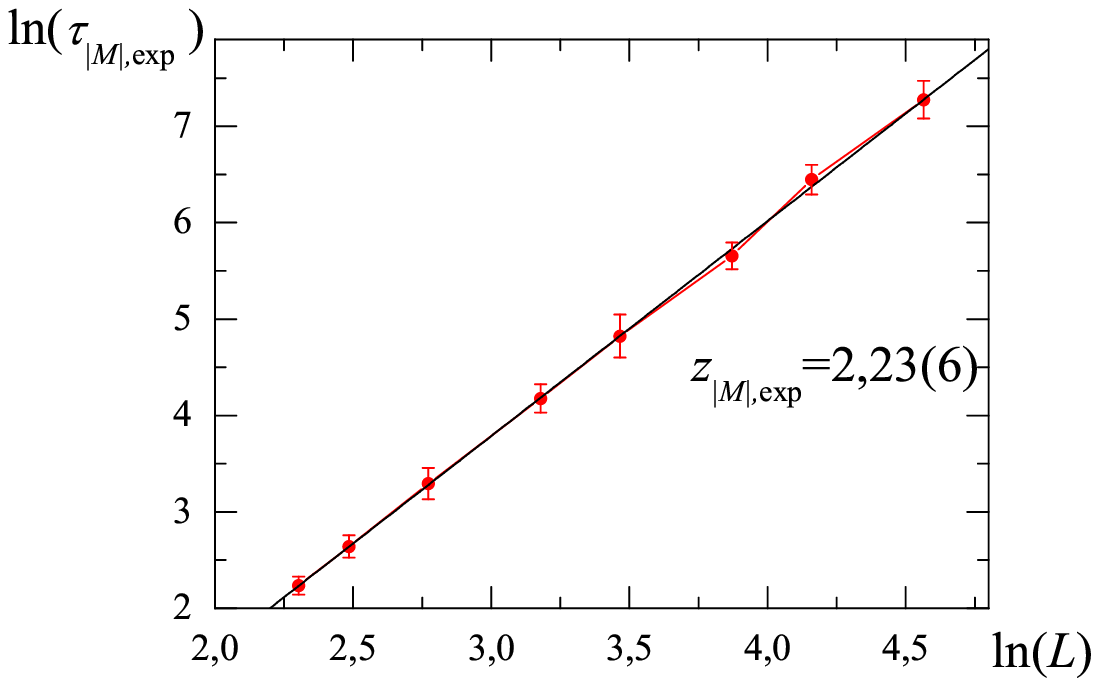}
\centerline{(e)\hspace{8cm}(f) } \caption{Integrated (left column)
and exponential (right column) autocorrelation times as functions
of $L$ for the Metropolis algorithm. a,b: energy autocorrelation,
$\tau_{E,{\rm int}}$, $\tau_{E,{\rm exp}}$; c,d: magnetization
autocorrelation, $\tau_{M,{\rm int}}$, $\tau_{M,{\rm exp}}$; e,f:
absolute value of magnetization autocorrelation, $\tau_{|M|,{\rm
int}}$, $\tau_{|M|,{\rm exp}}$. } \label{fig5}
\end{figure}

Having found autocorrelation times for different observables as
functions of lattice size one can extract via Eqs. (\ref{fff1}),
(\ref{fff2}) the values of dynamical exponents for each of the
algorithms considered. Let us start with the Metropolis algorithm.
Log-log plots for the integrated and exponential autocorrelation
times for $E$, $M$, and $|M|$ are shown in Fig.~\ref{fig5}. We
used a linear square interpolation to extrapolate the $\tau(L)$
dependencies to get the values of the exponents. To extrapolate
data obtained from the Metropolis algorithm all eight data points
were used, whereas for Swendsen-Wang and Wolff algorithms only
five last data points (the largest system sizes) were considered.

\begin{figure}[ht]
\epsfxsize=7cm \epsffile{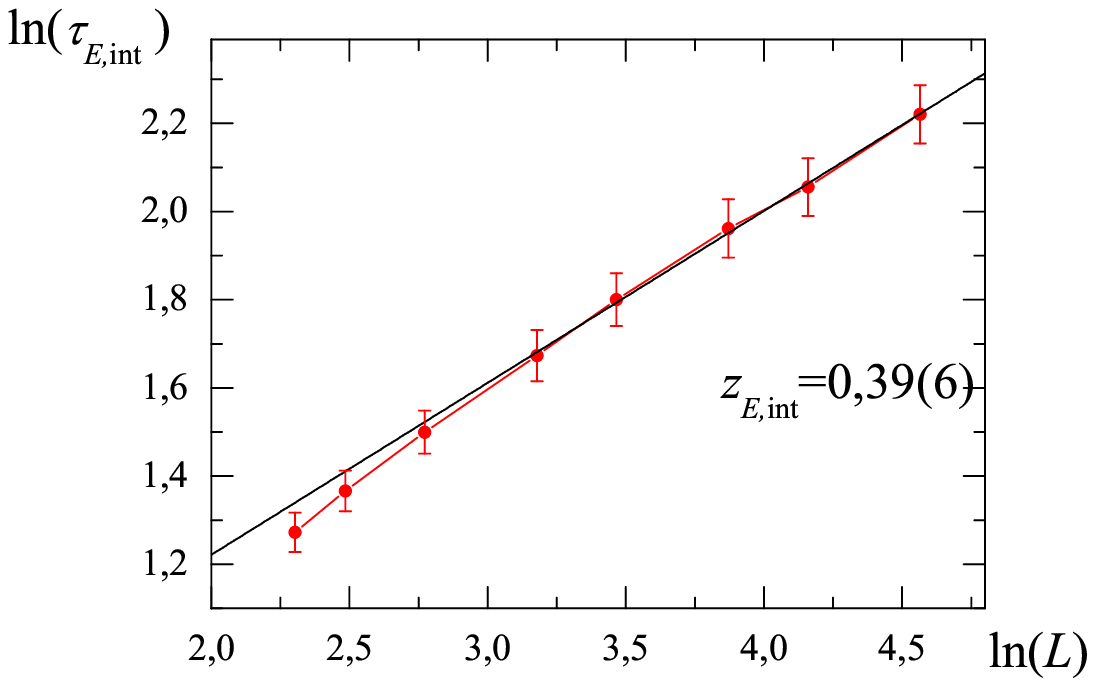} \epsfxsize=7cm
\epsffile{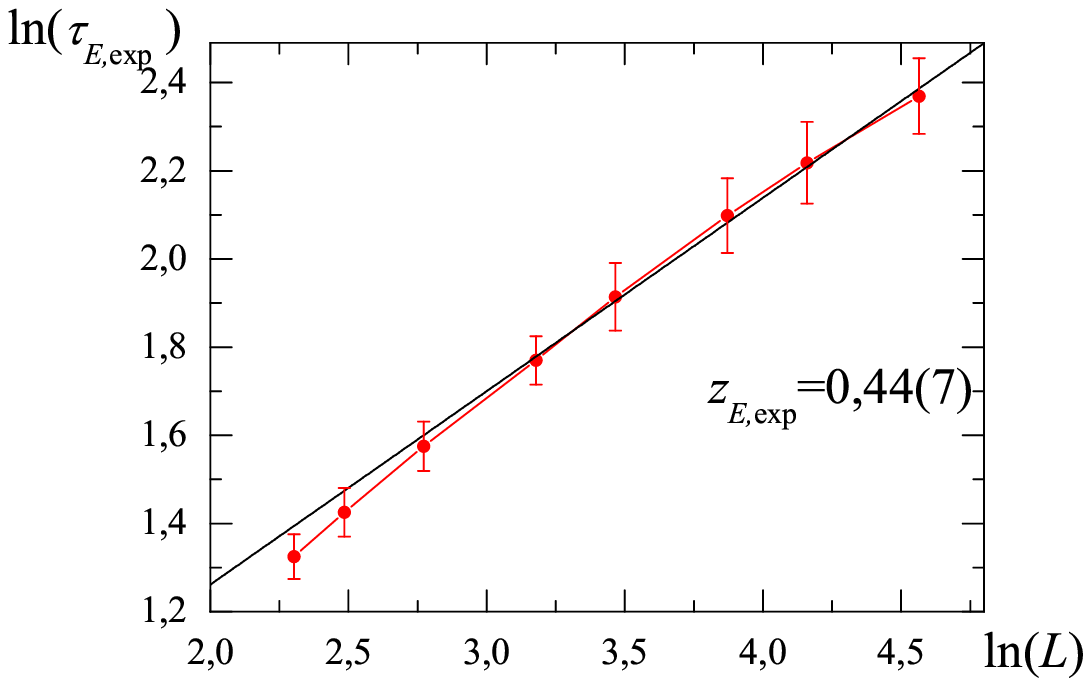} \centerline{(a)\hspace{8cm}(b) }
\epsfxsize=7cm \epsffile{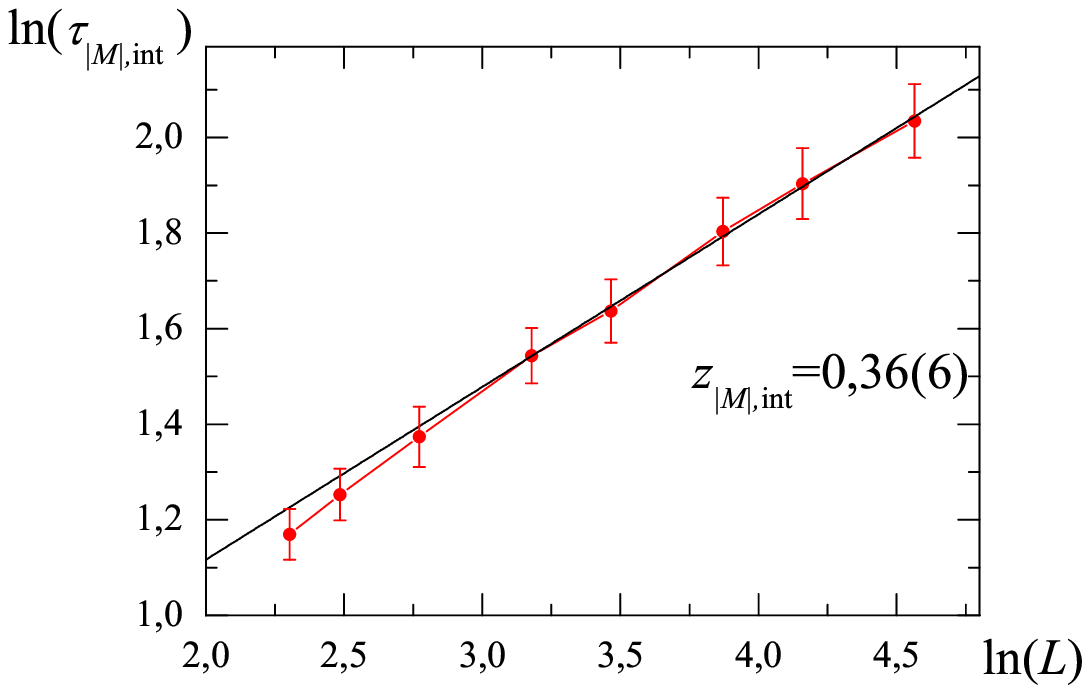} \epsfxsize=7cm
\epsffile{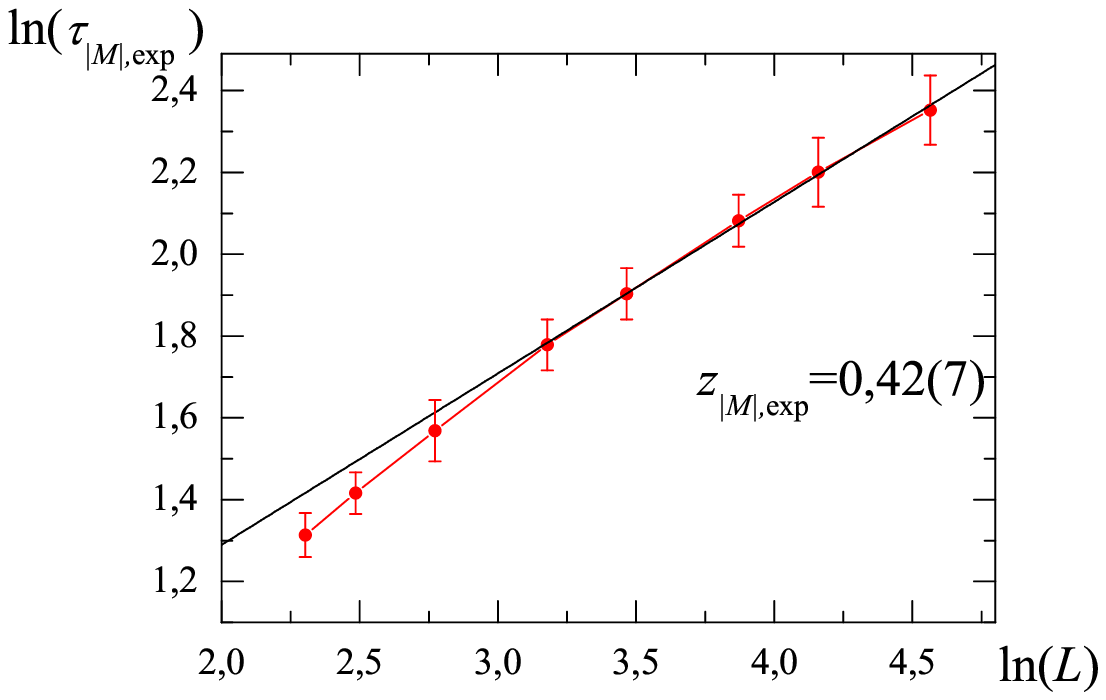} \centerline{(c)\hspace{8cm}(d) }
\caption{Integrated (left column) and exponential (right column)
autocorrelation times as functions of $L$ for the Swendsen-Wang
algorithm. a,b: energy autocorrelation, $\tau_{E,{\rm int}}$,
$\tau_{E,{\rm exp}}$; c,d: absolute value of magnetization
autocorrelation, $\tau_{|M|,{\rm int}}$, $\tau_{|M|,{\rm exp}}$.
\label{fig6} }
\end{figure}

\begin{figure}[ht]
\epsfxsize=7cm \epsffile{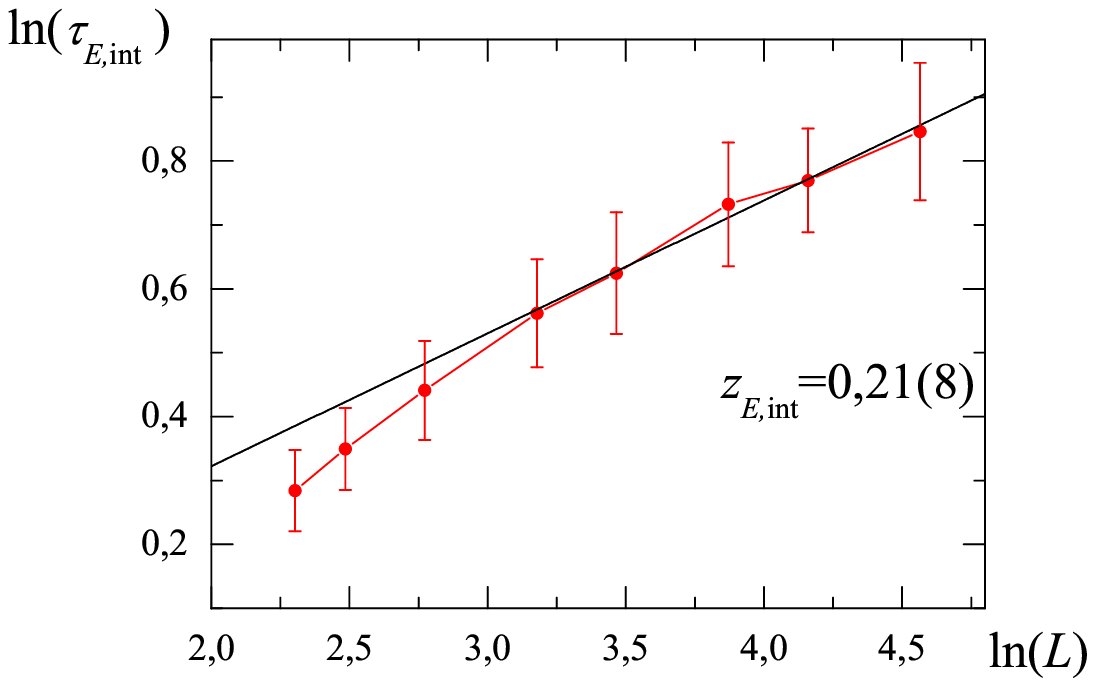} \epsfxsize=7cm
\epsffile{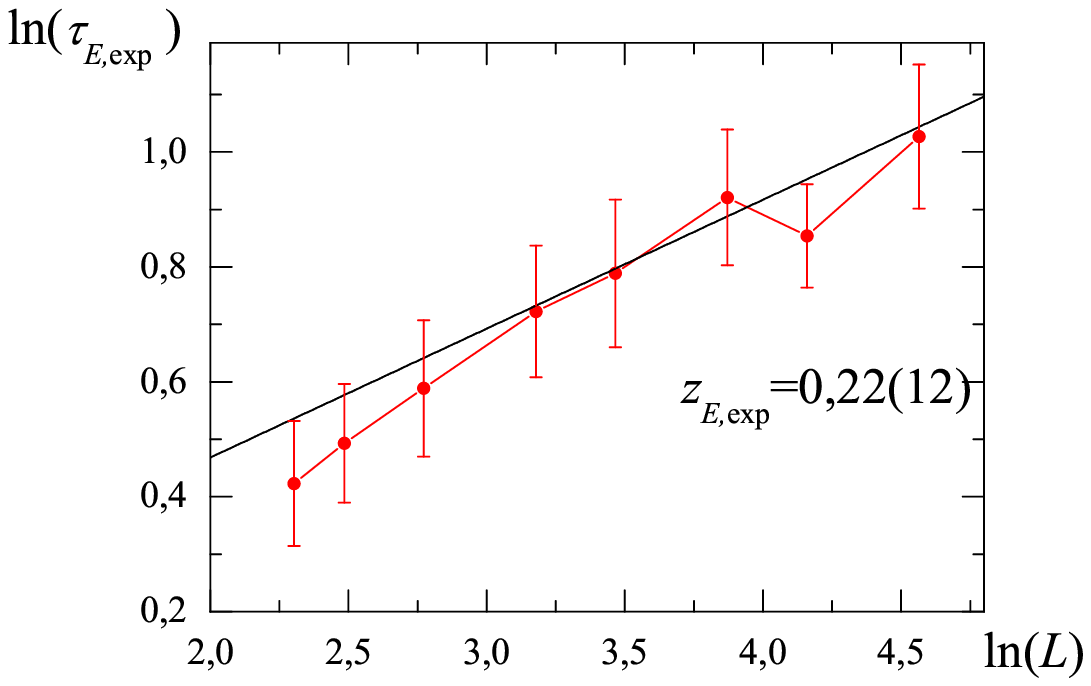} \centerline{(a)\hspace{8cm}(b) }
\epsfxsize=7 cm \centerline{\epsffile{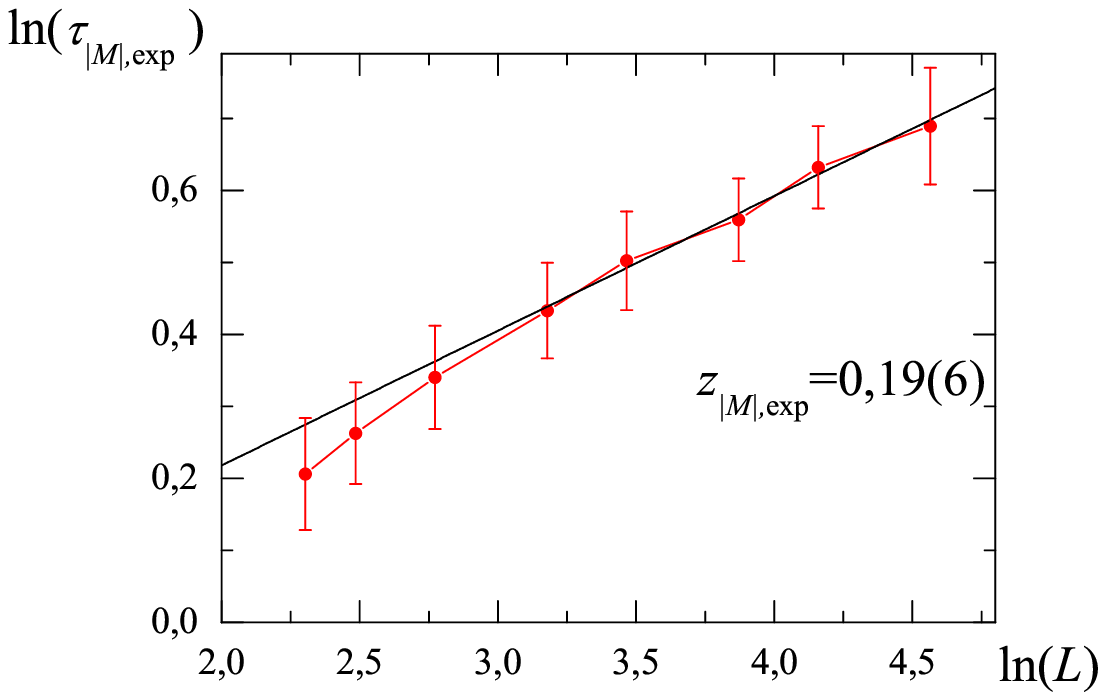}}
\centerline{(c)} \caption{ Autocorrelation times as functions of
$L$ for the Wolff algorithm. a,b: energy autocorrelation,
$\tau_{E,{\rm int}}$, $\tau_{E,{\rm exp}}$; c: absolute value of
magnetization autocorrelation, $\tau_{|M|,{\rm exp}}$.}
\label{fig7}
\end{figure}

Log-log plots for the autocorrelation time for cluster algorithms
are shown in Fig. \ref{fig6} (Swendsen-Wang algorithm) and Fig.
\ref{fig7} (Wolff algorithm). As noted above, not all
autocorrelation times are well-defined for the cluster algorithms.
In particular, for the Swendsen-Wang algorithm we were able to
define autocorrelations for $E$ and $|M|$ (and not for the
magnetization per site $M$), whereas for the Wolff algorithm only
$\tau_{E,{\rm int}}$, $\tau_{E,{\rm exp}}$, and $\tau_{|M|,{\rm
exp}}$ were well-defined. However, to estimate value of the
exponent $z_{E,{\rm exp}}$ we have to discard data for $L=64$. As
one can see from the plot in Fig.~\ref{fig7}b an appropriate data
point when included does not lead to a reasonable linear
approximation.

Numerical values of the exponents are given in the first line of
Table~\ref{tab6}. With the exception of the exponent for the
integrated energy autocorrelation time $\tau_{E,{\rm int}}$, the
rest of the exponents definitely are close to the value $z\simeq
2.2$. Comparing this value with the data of Table~\ref{tab1} one
sees, that it is in a reasonable agreement with the theoretical
estimates of
\cite{Grinstein77,Prudnikov92,Prudnikov98,Janssen95,Blavatska05}
as well as with the experimental result of \cite{Rosov92} and data
of MC simulations \cite{Heuer93,Prudnikov92a}. Note however, that
estimates from the out-of-equilibrium MC simulations give rather
different value $z\simeq 2.6$ \cite{Parisi99,Schehr05}.

\begin{table}[htb]
\caption{\label{tab6} RIM dynamical critical exponents for
different MC algorithms, obtained from data at the critical
temperature of the infinite system, Tabs.~\ref{tab4}-\ref{tab5}}
 \begin{center}
 {\small
\begin{tabular}{lllllll}
\hline & $z_{E,{\rm int}}$&$z_{E,{\rm exp}}$&$z_{|M|,{\rm
int}}$&$z_{|M|,{\rm exp}}$&$z_{M,{\rm int}}$&$z_{M,{\rm exp}}$\\
\hline
Metropolis&1.92(4)&2.16(4)&2.18(4)&2.23(6)&2.21(16)&2.29(19)\\
Swendsen-Wang&0.39(6)&0.44(7)&0.36(6)&0.42(7)&--&--\\
Wolff&0.21(8)&0.22(12)&--&0.19(6)&--&--\\ \hline
\end{tabular}
}
\end{center}
\end{table}

A rather striking feature of the dynamical exponents of cluster
algorithms is that dilution leads to decrease of the exponents, as
compared to the pure 3d Ising model. Indeed, the value for the
integrated dynamical critical exponent of the Swendsen-Wang
algorithm for the "energy like" observables recently calculated in
Ref. \cite{Ossola04} reads: $z_{{\mathcal E},{\rm
int}}=0.459(30)$. In the same study, the dynamical critical
exponent associated to the exponential autocorrelation time was
found to be $z_{\rm exp}=0.481$. Both values exceed those found by
us for the RIM, see the second line of Table~\ref{tab6}. Similar
tendency to the dilution induced decrease of the Swendsen-Wang
dynamical critical exponent was observed recently for the
random-bond 3d Ising model \cite{Berche04}. There, the value
$z=0.41$ for the bond concentration $p=0.7$ was found, which again
is smaller than its counterpart for the pure 3d Ising model
\cite{Ossola04}.

We re-analysed simulation data, considering them at the critical
temperature of a finite system of size $L$, $T_c(L)$. The latter
may be calculated in different ways, being defined by the maximum
of different observables. In Table~\ref{tab7}, we give the values
of RIM dynamical critical exponents at $T_c(L)$ obtained from the
maximum of magnetic susceptibility. The critical temperature was
taken from our previous study \cite{Ivaneyko05}. As one can see
comparing Tables \ref{tab6} and \ref{tab7}, the crossover effects
do not influence data essentially.

\begin{table}[htb]
\caption{RIM dynamical critical exponents for different MC
algorithms, calculated at the critical temperature of the finite
size system $T_c(L)$. \label{tab7}}
\small{\begin{center}
\begin{tabular}{lllllll}
\hline & $z_{E,{\rm int}}$&$z_{E,{\rm exp}}$&$z_{|M|,{\rm
int}}$&$z_{|M|,{\rm exp}}$&$z_{M,{\rm int}}$&$z_{M,{\rm exp}}$\\
\hline
Metropolis&1.99(3)&2.22(3)&2.19(5)&2.22(7)&2.18(12)&2.23(16)\\
Swendsen-Wang&0.35(5)&0.39(6)&0.32(6)&0.40(6)&--&--\\
Wolff&0.16(8)&0.16(9)&--&0.14(5)&--&--\\ \hline
\end{tabular}
\end{center}
}
\end{table}

One more question worth discussing is whether the relations
between cluster algorithms dynamical exponents and the static
exponents observed for the pure systems \cite{Coddington92,Li89}
hold for the diluted ones. Indeed, for the pure Ising model, the
Coddington-Ballie conjecture holds \cite{Coddington92}, stating
that the Swendsen-Wang dynamical critical exponent $z^{\rm SW
}_{E,{\rm int}}$ is defined via static critical exponents for
magnetization and correlation length:
\begin{equation}\label{sw}
z^{\rm SW}_{E,{\rm int}}=\beta/\nu.
\end{equation}
It is worth here to note, that whereas the static critical
exponents for RIM numerically differ from those of the pure 3d
Ising model (cf. theoretical RG estimates $\beta=0.349(5)$ and
$\nu=0.678(10)$  \cite{Pelissetto00} for RIM with
$\beta=0.3258(14)$ and $\nu=0.6304(13)$
 \cite{Guida98} for pure 3d Ising model) their
relation remains almost unchanged. For the numbers given above,
$\beta/\nu=0.515(15)$ for RIM and $\beta/\nu=0.517(3)$ for the 3d
Ising model. Therefore, a change in the value of the Swendsen-Wang
dynamic critical exponent  upon dilution serves as an evidence
that the relation (\ref{sw}) does not hold for RIM.

Another empirical relation found in Ref. \cite{Coddington92} for
the pure Ising model connects the dynamical critical exponent of
the Wolff algorithm $z^{\rm W}_{E,{\rm int}}$ with the static
ones:
\begin{equation}\label{w}
z^{\rm W}_{E,{\rm int}}=\alpha/\nu.
\end{equation}
As far as $\alpha<\beta$ for the 3d Ising model, comparison of
Eqs. (\ref{sw}) and (\ref{w}) leads to the inequality:
\begin{equation}\label{ineq}
z^{\rm W}_{E,{\rm int}}\, < \, z^{\rm SW}_{E,{\rm int}}.
\end{equation}
Eq. (\ref{w}) does not hold for the diluted systems (where the
heat capacity critical exponent is negative). However, the
inequality (\ref{ineq}) still holds, as one can see, comparing
data of Table \ref{tab6} for the Wolff and Swendsen-Wang
algorithms. Again, the value of dynamical critical exponent for
the diluted system is smaller than its counterpart for the pure
one. Wolff algorithm dynamical critical exponent of the 3d Ising
model found in different simulations read: $z_{E, {\rm int}}=$
0.28(2) \cite{Wolff89a}; 0.44(10) \cite{Tamayo90}; 0.33(1)
\cite{Coddington92}, all numbers exceeding those given for the
Wolff case of RIM in Table~\ref{tab6}.

\section{Conclusions and outlook}
\label{VII}

In this paper, we have studied dynamical critical behaviour of the
3d random-site Ising model (RIM) originated from different MC
algorithms. We considered the local single-spin Metropolis
algorithms as well as Swendsen-Wang and Wolff cluster algorithms.
Giving origin to an equivalent static critical behaviour, all
three algorithms correspond to different forms of dynamics. A
comparison of numerical characteristics of Metropolis (local) and
Swendsen-Wang and Wolff (cluster) dynamics for RIM was achieved by
calculation of the integrated and exponential autocorrelation
times for RIM energy and magnetization.

The local update Metropolis algorithm corresponds to the  pure
relaxational single-spin dynamics with non-conserved order
parameter and finds its theoretical description as the model A
critical dynamics \cite{Hohenberg77}. There exist RG analysis of
critical dynamics for the RIM with such type of relaxation
\cite{Grinstein77,Prudnikov92,Prudnikov98,Janssen95,Oerding95,Blavatska05}.
It assumes, however, a single dynamical critical exponent $z$ for
the relaxation times of different observables. Although the
perturbation theory series are divergent and only the lowest
non-trivial order calculations have been performed so far, being
resummed appropriately, all available theoretical data are
coherent with an estimate $z \simeq 2.2(1)$ (see Table~\ref{tab1}
of our paper). This is further supported by the latest
experimental observation we are aware about, $z=2.18(10)$
\cite{Rosov92}. Whereas initial MC simulations gave estimates of
$z$ along with the above value \cite{Heuer93,Prudnikov92a}, recent
simulations of Refs. \cite{Parisi99,Schehr05} favour an estimate
$z \simeq 2.6(1)$. Our results for the values of RIM dynamical
critical exponent for a local dynamics are summarized in the first
line of Table~\ref{tab6}. The essential features of the discussion
remain unchanged, when one considers calculations at the critical
temperature of the finite size system, $T_c(L)$, Table~\ref{tab7}.
Except of the exponent for the energy integrated autocorrelation
time, $z_{E,{\rm int}}$, our data supports an estimate $z \simeq
2.2(1)$ within different error bars, corresponding to different
observables measured during simulations. The discrepancy between
our estimates an those of Refs. \cite{Parisi99,Schehr05} may be
caused by the fact, that the latter have defined scaling exponents
for the out-of-equilibrium short-time dynamics.

The Swendsen-Wang and Wolff algorithms give rise to the dynamics
of spin clusters, which differs from the local one. Even for the
pure 3d Ising model there is no field theory describing such
dynamics. However, there exist estimates, relating dynamical
critical exponent of the cluster algorithms to the static
exponents. Besides Eqs. (\ref{sw}), (\ref{w}), the following
inequality has been proven for the energy-like integrated and
exponential  autocorrelation time critical exponents of
Swendsen-Wang algorithm \cite{Li89}:
\begin{equation}\label{liineq}
z^{\rm SW}_{{\mathcal E},{\rm int}}, z^{\rm SW}_{\rm exp} \geq
\alpha/\nu.
\end{equation}
Eqs. (\ref{sw}), (\ref{w}), (\ref{liineq}) hold for the pure Ising
model. In particular, Eq. (\ref{liineq}) leads to the conclusion,
that systems with a positive specific heat exponent $\alpha$ must
display critical slowing down. In absence of such inequality for
the diluted system, our results for the critical dynamics of RIM
for cluster algorithms prove that the critical slowing down is
present in diluted systems as well. However, a striking feature of
our estimates (second and third lines of table~\ref{tab6}) is that
they suggest that dilution leads to decrease of the dynamical
critical exponent for the cluster algorithms. This phenomena is
quite opposite to the local dynamics, where dilution enhances
critical slowing down. The values of the exponents describing
relaxation of different observables differ numerically, being
however close to each other. Nevertheless, on this stage it is
impossible to exclude that the difference is not only due to the
crossover phenomena \cite{note2}.

\section*{Acknowledgements}

We acknowledge useful discussions with Christophe Chatelain, Maxym
Dudka, Reinhard Folk, and Wolfhard Janke. Work of Yu.H. was
supported in part by the Austrian Fonds zur F\"orderung der
wissenschaftlichen Forschung, project No. 16574 PHY.

\end{document}